\documentclass{iaus}
\usepackage{graphicx}
\usepackage[square, comma, sort&compress]{natbib}

\title[M31 Major Merger, the LMC Pogenitor] 
{Could M31 come from a major merger and eject the LMC away?}

\author[S. Fouquet, F. Hammer, Y. B. Yang, J. L. Wang, M. Puech, H. Flores]   
{S. Fouquet$^1$, F. Hammer$^1$, Y. B. Yang$^2$, J. L. Wang$^{1,2}$, M. Puech$^1$ \& H. Flores$^1$}

\affiliation{
$^1$Laboratoire GEPI, Observatoire de Paris, CNRS-UMR8111,
\\Univ. Paris-Diderot 5 place Jules Janssen, 92195 Meudon France
\\ email: {\tt francois.hammer@obspm.fr} \\[\affilskip]
$^2$NAOC, Chinese Academy of Sciences,
\\A20 Datun Road, 100012 Beijing, China
}

\pubyear{2010}
\volume{277}  
\pagerange{1--4}
\jname{Tracing the ancestry of galaxies on the land of our ancestors}
\editors{A.C. Editor, B.D. Editor \& C.E. Editor, eds.}

\begin{document}

\maketitle

\begin{abstract}

We investigated a scenario in which M31 could be the remnant of a major merger and at the origin of the LMC. Galaxy merger simulations were run in order to reproduce some M31 properties. We succeeded in reproducing some of the most important M31 large-scale features like the thick disk or the polar ring, and gave a possible explanation for the formation of the Giant Stream. We also found that the LMC could be expelled by this high energetic phenomenon.

\keywords{galaxies: formation - galaxies: Local Group - galaxies: Magellanic Cloud}
\end{abstract}

\firstsection 
\section{Introduction}

Several observations suggest that the Andromeda galaxy could be the remnant of a major merger. M31 has a robustly classical and not pseudo bulge, which is an argument in favor of such a merger scenario, according to \citet{Kormendy2010}. \citet{VdB2005} also argued that due to the high metallicity of the M31 halo and the $r^{1/4}$ luminosity profile of the bulge, M31 is likely the result of a merger of two massive metal-rich ancestors. Conversely, there are several difficulties in reproducing the Giant Stream by a recent minor merger, according to \citet{Font2008}. For instance, they do not succeed in reproducing the metal-poor ([Fe/H]$<-2$) stellar population. In addition, a recent merger should have produced a fraction of young stars that is not detected. A last argument is given by \citet{Ferguson2005}. Around the Andromeda galaxy, several other features like shelves or clumps have been detected using deep exposures with HST/ACS. \citet{Ferguson2005} found a moderate range of age and metallicity in these structures, which suggest a simultaneous formation process.

Although a major merger seems to be a likely event in the M31 lifetime, other mechanisms are likely to occur in parallel such as minor mergers or gas accretion, and be at the origin of some M31 sub-structures.

\section{Dating a possible major merger}

\begin{figure}[!h]
  \begin{center}
    \includegraphics[width=0.85\linewidth]{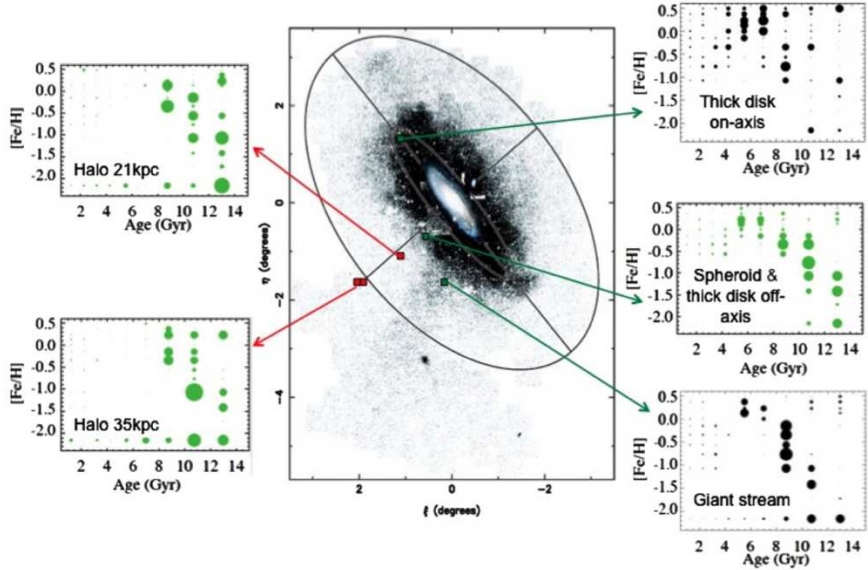} 
    \caption{Chronological sketch of the structures surrounding M31. In the central panel (reproduced from \citet{Ibata2005}, the large and thick rotating disk is a vast flattened structure with a major axis of about 4$^{\circ}$. Squares represent fields observed by \citet{brown06, brown07, brown08}, and are linked to their measurements by arrows. Figure extracted from \citet{Hammer2010}.}
    \label{fig1}
  \end{center}
\end{figure}

We propose, from Figure \ref{fig1}, a chronological history of the different structures in M31, as this figure can be used as a clock for determining the occurrence of merger phases as follows. The star formation history of a merging system is enhanced during the first passage until the fusion and then at the fusion itself \citep[see][]{cox08}. During such a major merger event, most of the gas and stars in the remnant outskirts is deposited by tidal tails formed during the first passage and later during the fusion of the cores. A few hundred billion years after its formation, the tidal tail dilutes, provoking a natural quenching of the residual star formation \citep[see][]{wetzstein07}. Thus, the age of the material brought by tidal tails provides, with a relatively small delay, possible dates for both the first passage and fusion times. In the following, we assumed that the first passage occurred from 8.5 to 9 Gyr ago, and that the corresponding tidal tails are responsible for the halo enrichment seen in the 21 and 35 kpc fields, without significant star formation more recent than 8.5 Gyr. The thick disk has a star formation history comparable to that of the Giant Stream and is also generated by material returning to the galaxy mostly from tidal tails generated at the fusion. Because their youngest significant population of stars has ages of 5.5 Gyr, the delay between the first passage and fusion ranges between 3 and 3.5 Gyr. This could be accommodated for by relatively large impact parameters (20-30 kpc).

\section{Simulation constraints and results}

The GADGET2 hydrodynamical code \citep{springel05} supplemented by star formation, feedback and cooling prescriptions \citep{cox08} was used. For simulating large scale aspects (thin and thick disk, B/T ratio, 10 kpc ring), low resolution are enough, typically $\sim 10^5$ particles. However, we had to increase the number of particles for reproducing the Giant Stream, i.e, more than $5 \times 10^5$. Other structures could not be simulated with such a resolution (like the double nucleus) because they require a much better spatial resolution. In addition, some features can be formed by minor mergers, or other mechanisms completely independent of the major merger process assessed here.

\begin{figure}[!ht]
  \begin{center}
    \includegraphics[width=0.85\linewidth]{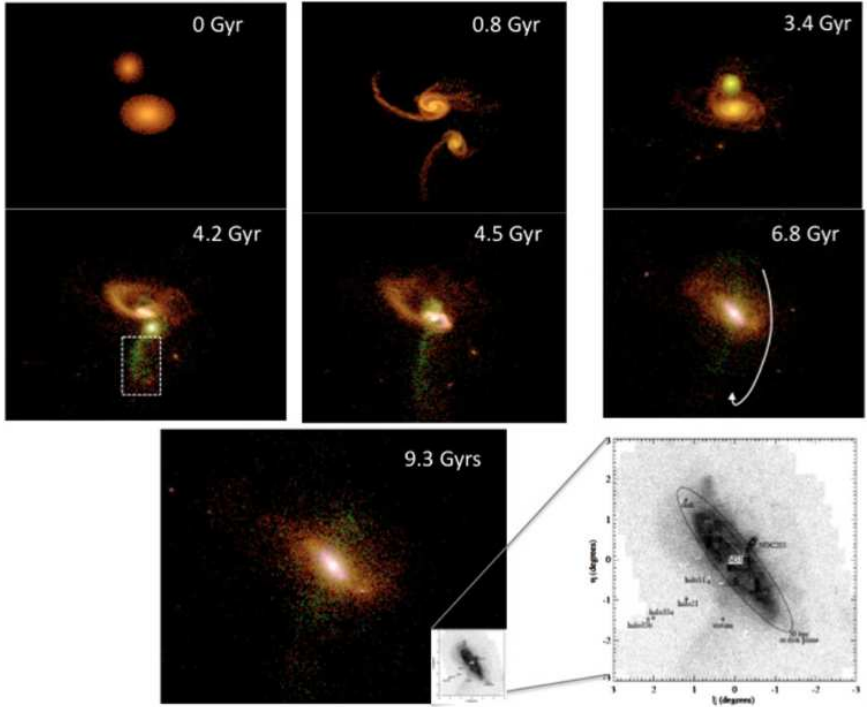}
    \caption{Different phases of a 3:1 major merger for M31 (rpericenter = 24 kpc, Gal1 incy = 70, Gal2 incy = −110). The simulation starts 9.3 Gyr ago (T = 0, z = 1.5), and the first passage occurs 0.7 Gyr later. Then new stars are forming (green color), especially in the secondary galaxy, until the fusion, which occurs at 4.5 Gyr. The elapsed time between the first passage and fusion is 3.8 Gyr, as the pericenter radius is large. During the second passage (T = 4.2 Gyr) a tidal tail containing many newly formed intermediate-age stars (green dots) is formed. Later on this material returns to the galaxy forming the Giant Stream, enriched with stars formed from 5 to 8 Gyr ago. The resulting galaxy in the last panel (T = 9.3 Gyr) is compared with the inserted M31 image at the same scale (from \citet{Ibata2005}, see an enlarged view of this insert on bottom right). Figure extracted from \citet{Hammer2010}.}
    \label{fig2}
  \end{center}
\end{figure}

Observations strongly constrain the parameters of the simulations. The first constraints are the rotation velocity curve of the thin disk and the B/T ratio. According to \citet{hopkins09}, a mass ratio from $\sim$3:1 to $\sim$2:1 is required to produce a remnant bulge with B/T $\sim$ 0.3. Another constraint is the total baryonic mass for the progenitors, which must be close to the M31 baryonic mass, i.e., $1.1 \times 10^{11}$ M$_{\odot}$. A prograde-retrograde orientation for the spin axis was chosen, which is favourable to rebuild disk \citep[see][]{hopkins08}.

Considering the gas fraction, reforming an Sb-like thin disk demands high gas fraction just before the fusion, i.e more than 50\%. As a result, low star formation and gas-rich progenitors, with more than 65\% of gas, were used in order to keep enough gas before the fusion. These assumptions have theoretical grounds. It has been estimated that many high-z galaxies are gas rich, hence M31 progenitors should have a high gas fraction. Low star formation could be caused either by higher feedback efficiency in primordial medium, or because the gas in the progenitors is less concentrated than in present-day spirals as it is in present-day low-surface brightness galaxies, or because cooling is less efficient in a relatively pristine medium, or a combination of all these factors. More generally, the expected increase of the gas metal abundance (expected to be slow before the fusion but very efficient during the fusion) may help to increase the molecular gas fraction, the optical depth of the gas, and the radiation pressure effects, all contributing to a change in the star formation history during the interaction (T. J. Cox 2010, private communication).

Another structure to be reproduced was the 10 kpc ring, which suggests a polar orbit. The last feature, the Giant Stream, is caused by particles coming back from the tidal tail formed just before fusion. Figure \ref{fig2} describes the formation of such a structure that is aligned along the trajectory of the satellite which falls into the mass center at the fusion, 4.5 Gyr after the beginning of the simulation. We verified that, within the family of orbits we chose, the strength of the tidal tail (see Figure \ref{fig2}, panel at 4.2 Gyr) depends on the inclination of the progenitors relative to the orbital angular momentum.

\section{LMC could be ejected from Andromeda}

The LMC has a particular status in the Local group: it is the only dwarf irregular within 250 kpc of the MW. With new velocity measures, close to 380 km/s \citep{Kalli06}, and a distance from the Milky Way equal to 50 kpc, the idea that the LMC go past for the first time around the Milky Way was revived. By tracing back the LMC, it goes back far away from the MW. It results many configurations for which the LMC was close to M31 in the past; and could have been formed as a tidal dwarf during the merger event. The radial velocity between the MW and M31, $\sim$ 136 km/s, is known but the tangential one is not measured for the moment because it is too small. To test this idea, simulations were run with different values for the M31 tangential velocity and we were looked for which one the LMC would go back to M31 \citep{YY2010}.

For a M31 proper motion equal to $ \mu_W = -62 \, \mu \mathrm{as.yr}^{-1}$, $\mu_N = -25 \, \mu \mathrm{as.yr}^{-1}$ in the equatorial system, the resulting tangential velocity is $106$ km/s and the encounter between the LMC and M31 occurred 5.5 Gyr ago.

\end{document}